\documentclass[%
reprint,
superscriptaddress,
amsmath,amssymb,
aps,
prl,
floatfix,
]{revtex4-2}

\usepackage{braket}
\usepackage{graphicx}% Include figure files
\usepackage{dcolumn}% Align table columns on decimal point
\usepackage{bm}% bold math
\usepackage{hypernat}
\usepackage{hyperref}% add hypertext capabilities
\usepackage{cleveref}
\usepackage[skip=1pt, indent=25pt]{parskip}
\usepackage{ulem}
\hypersetup{
    colorlinks,
    citecolor=black,
    filecolor=black,
    linkcolor=black,
    urlcolor=black,
    pdfstartview={FitH},
    pdftitle={Germanium Gatemon},
    pdfauthor={Sági Olivér}
}

\begin{document}

\preprint{Second draft}

\title{A gate tunable transmon qubit in planar Ge}%\\with Forced Linebreak}% Force line breaks with \\

\author{Oliver Sagi}
\email{oliver.sagi@ista.ac.at}
\affiliation{Institute of Science and Technology Austria, Am Campus 1, 3400 Klosterneuburg, Austria}
\author{Alessandro Crippa}
\affiliation{NEST, Istituto Nanoscienze-CNR and Scuola Normale Superiore, I-56127 Pisa, Italy}
\author{Marco Valentini}
\affiliation{Institute of Science and Technology Austria, Am Campus 1, 3400 Klosterneuburg, Austria}
\author{Marian Janik}
\affiliation{Institute of Science and Technology Austria, Am Campus 1, 3400 Klosterneuburg, Austria}
\author{Levon Baghumyan}
\affiliation{Institute of Science and Technology Austria, Am Campus 1, 3400 Klosterneuburg, Austria}
\author{Giorgio Fabris}
\affiliation{Institute of Science and Technology Austria, Am Campus 1, 3400 Klosterneuburg, Austria}
\author{Lucky Kapoor}
\affiliation{Institute of Science and Technology Austria, Am Campus 1, 3400 Klosterneuburg, Austria}
\author{Farid Hassani}
\affiliation{Institute of Science and Technology Austria, Am Campus 1, 3400 Klosterneuburg, Austria}
\author{Johannes Fink}
\affiliation{Institute of Science and Technology Austria, Am Campus 1, 3400 Klosterneuburg, Austria}
\author{Stefano Calcaterra}
\affiliation{L-NESS, Physics Department, Politecnico di Milano, via Anzani 42, 22100, Como, Italy}
\author{Daniel Chrastina}
\affiliation{L-NESS, Physics Department, Politecnico di Milano, via Anzani 42, 22100, Como, Italy}
\author{Giovanni Isella}
\affiliation{L-NESS, Physics Department, Politecnico di Milano, via Anzani 42, 22100, Como, Italy}
\author{Georgios Katsaros}
\affiliation{Institute of Science and Technology Austria, Am Campus 1, 3400 Klosterneuburg, Austria}

\date{\today}% It is always \today, today,

\begin{abstract}

Gate-tunable transmons (gatemons) employing semiconductor Josephson junctions have recently %\textcolor{blue}{have recently} 
emerged as building blocks for hybrid quantum circuits.
In this study, we present a gatemon fabricated in planar Germanium. We induce superconductivity in a two-dimensional hole gas by evaporating aluminum atop a thin spacer, which separates the superconductor from the Ge quantum well. The Josephson junction is then integrated into an Xmon circuit and capacitively coupled to a transmission line resonator. We showcase the qubit tunability in a broad frequency range with resonator and two-tone spectroscopy. Time-domain characterizations reveal energy relaxation and coherence times up to $75\, \textrm{ns}$. Our results, combined with the recent advances in the spin qubit field, pave the way towards novel hybrid and protected qubits in a group IV, CMOS-compatible material.

\end{abstract}

\maketitle

%\tableofcontents

\section{Introduction}
%\showthe\textwidth

% Hybrid quantum circuits combine two or more physical systems to harness the advantages and strengths to bring new quantum technologies.\\
%Hybrid quantum circuits, which integrate two or more physical paradigms in the same platform, have gained increasing interest as they can lead to devices with novel functionalities and new phenomenology~\cite{RevModPhys.85.623, 10.1063/5.0024124}.
%\textcolor{green}{what do you mean by new phenomenology? Is there a difference between functionality and phenomenology here?}

Hybrid quantum circuits interface different physical paradigms in the same platform, leading to devices with novel functionalities~\cite{RevModPhys.85.623, 10.1063/5.0024124}. 
%In solid-state physics, this materializes in novel nanodevices unraveling unexplored phenomenologies. 
Specifically, hybrid qubits combining superconductors and semiconductors merge the maturity of superconducting quantum circuits with the inherent tunability of semiconductors.
Gate-tunable transmons (gatemons)~\cite{de2015realization,PhysRevLett.115.127001,Casparis2018}, parity-protected qubits~\cite{PhysRevLett.125.056801}, and Andreev spin qubits (ASQs)~\cite{doi:10.1126/science.abf0345,Pita-Vidal2023} are representative examples of recent achievements.
In the development of hybrid qubits on novel material platforms, demonstrating coherent interaction between a microwave resonator and a Josephson junction is a paramount step. 
Experimentally, the most direct approach to achieve this interaction is through capacitive coupling between a resonator and a gatemon circuit~\cite{Wang2019,Chiu2020,PhysRevApplied.15.064050,Schmitt2022}. 
%\textcolor{red}{Why is the simplest? Either we explain or we change word. For me a gatemon circuit includes already the resonator, do you agree?}.
This coupling enables dispersive readout~\cite{RevModPhys.93.025005} and, when combined with coherent control, serves as the foundation for progressing towards advanced hybrid circuits.
%including ASQs.

A prominent example is ASQs, which combine the advantages of both superconducting and semiconductor spin qubit platforms~\cite{Pita-Vidal2023,doi:10.1126/science.abf0345,cheung2023photonmediated,pitavidal2023strong,PhysRevB.106.115411}.
%Integrating gatemons in ASQ circuits also loosens up readout requirements by allowing in-situ frequency tuning and capacitive coupling to resonators~\cite{PRXQuantum.3.030311}. %\textcolor{red}{This time there's a reference that help, but either we stay more general (taking away the coherent sentence, or we explain shortly what we mean}.
%finds utility in novel ASQ circuits as well~\cite{PRXQuantum.3.030311}. 
%\textcolor{green}{I would remove from: , including ASQs}
%textcolor{red}{ASQs are now in the spotlight due to their promise of combining the advantages of both superconducting and semiconductor spin qubit platforms~\cite{Pita-Vidal2023,doi:10.1126/science.abf0345,cheung2023photonmediated,pitavidal2023strong,PhysRevB.106.115411}. - In the second sentence, this concept has already been presented. Do we really need this sentence here?} \textcolor{green}{A prominent example is ASQs which combine the advantages of both superconducting and semiconductor spin qubit platforms~\cite{Pita-Vidal2023,doi:10.1126/science.abf0345,cheung2023photonmediated,pitavidal2023strong,PhysRevB.106.115411} }
Semiconductor spin qubits offer low footprint and high anharmonicity, but implementing long-range coupling over many qubits remains challenging, though remarkable advances have been recently demonstrated~\cite{dijkema2023twoqubit,Borjans2020,PhysRevX.12.021026}. 
Superconducting transmon circuits provide well-mastered control and readout via microwave signals at the expense of large footprints and lower anharmonicity, posing challenges to scalability and fast operations. 
Finding a suitable platform to merge the two systems is a formidable task. 
It requires a low microwave-loss substrate (unless flip-chip technology is used~\cite{hinderling2024direct}), transparent semiconductor-superconductor interfaces, and material free from nuclear spins for optimal spin qubit operation. 

In the field of semiconductor spin qubits, Si is an attractive material choice as a result of the mature complementary metal-oxide-semiconductor (CMOS) technology~\cite{vinet2018towards} and isotopic purification~\cite{RevModPhys.95.025003}.
Moreover, Si-based platforms have demonstrated the capability of hosting microwave resonators coupled to quantum dots ~\cite{Mi2018,Yu2023,doi:10.1126/science.aar4054, Borjans2020, PhysRevX.12.021026}.
However, the absence of Fermi level pinning makes it challenging to proximitize silicon without using doping or annealing techniques~\cite{Bustarret2006,vethaak2022silicidebased}.
%\sout{Still, the superconductive proximity effect on Si has not yet been demonstrated.} 
III-V semiconductor compounds, on the other hand, have emerged as the natural solution for hybrid devices due to the high-quality epitaxial aluminum (Al) growth, yielding a hard gap (i.e. free of subgap states) and a transparent interface~\cite{PhysRevB.93.155402,Krogstrup2015,PhysRevApplied.7.034029,Lutchyn2018}. 
Nevertheless, the non-zero nuclear spin limits the spin qubit coherence to $\sim 10\,\textrm{ns}$~\cite{Nadj-Perge2010}. 
%To overcome the limitations of the former material systems we introduce a new platform that can potentially mitigate the problems above. 
%Here, we present a platform based on Ge heterostructure where an Al layer induces superconductivity evaporated on top of the buffer layer. 
Ge, and in particular Ge/SiGe heterostructures, have shown great potential for highly-coherent spin qubits and gate-tunable hybrid Josephson junctions~\cite{Hendrickx2018,Jirovec2021,hendrickx2023sweetspot,wang2024operating,doi:10.1021/acs.nanolett.8b04275,PhysRevResearch.3.L022005,Valentini2024,Scappucci2021}, %\textcolor{red}{We should cite first the spin qubit papers in chronological order, then the hybrid works in chronological order and finally the Ge review.}
%\textcolor{blue}{here we should cite the Hendrickx nature com 2018, Kushagra's paper, https://pubs.acs.org/doi/10.1021/acs.nanolett.8b04275, and our recent Nature Com. And I suggest we put at the end of this sentence all the references also for the spin part}. 
thus allowing the integration of semiconducting qubits with superconducting qubits on the same substrate.
\begin{figure*}[!htbp]
\includegraphics[width=\textwidth]{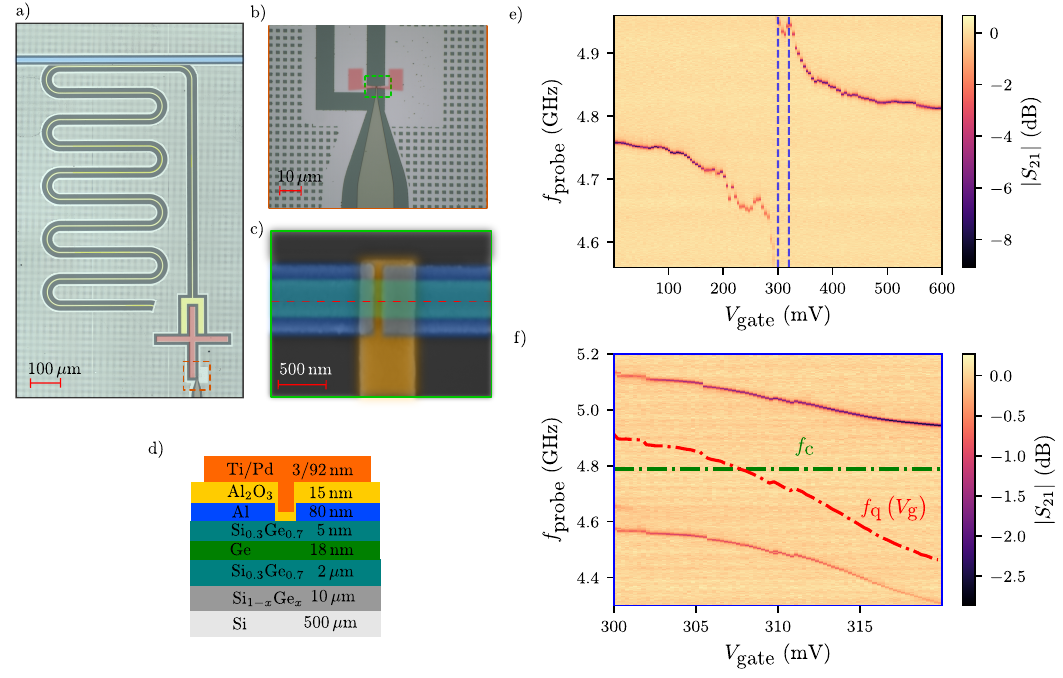}% Here is how to import EPS art
\caption{\label{fig:device} Overview of the complete Ge gatemon device and resonator spectroscopy measured with a Vector Network Analyzer. 
a) Optical microscope image showcasing the entire device.
A $\lambda/4$  notch-type coplanar waveguide resonator (yellow) is capacitively coupled to the cross-shaped qubit island (red) alongside a transmission line (blue) utilized for readout.
The island is shunted to the ground through a gate-tunable semiconductor Josephson junction. 
b)  Close-up view of the U-shaped mesa (highlighted in red).  
c) False-colored SEM image of the junction, depicting the junction where the evaporated aluminum (blue) extends over the mesa (light green) to proximitize the whole mesa.
The gate line (orange) is intentionally extended on the right side to increase capacitance to the ground. 
d) Cross-section of the wafer stack along the red-dashed line in Fig.~\ref{fig:device}c. 
e) Gate-dependence of the normalized resonator transmission after background correction.
f) Close-up view of the avoided crossing after background correction and boxcar averaging with a window size of 8 points. The dashed green line represents the bare resonator frequency, while the red dashed line indicates the qubit frequency extracted from the two peaks according to Eq.~\ref{eq:two}.} 
%\textcolor{red}{In the main text you say that you used the other equation}}
\end{figure*}

%%% Figure 1 %%%%

 Ge gatemons within Ge/Si core/shell nanowires have been recently realized~\cite{Zhuo2023,zheng2023coherent}. 
%However, scaling up to s enabling many-qubit connectivity~\cite{PhysRevB.106.115411,pitavidal2023strong} within one-dimensional platforms remains challenging.
However, while such CVD-grown core/shell wires have demonstrated ultrafast spin qubits~\cite{carballido2024qubit,Froning2021} they face challenges in terms of spin dephasing times and scale-up. %~\cite{PhysRevB.106.115411,pitavidal2023strong}.
To circumvent both issues, we build on the success of two-dimensional implementations in InAs/InGaAs heterostructures~\cite{Casparis2018,strickland2023characterizing}
and realize a gatemon based on a Ge/SiGe heterostructure where superconductivity is induced into the Ge hole gas by proximity from an Al layer evaporated on top of the SiGe spacer.
% \sout{However, the implementation of devices enabling many-qubit connectivity~\cite{PhysRevB.106.115411,pitavidal2023strong} within one-dimensional platforms remains strongly desirable in the prospect of scalability.}
% \sout{To address these challenges, we build on the success of two-dimensional implementations in InAs/InGaAs heterostructures~\cite{Casparis2018,strickland2023characterizing}.
% Here, we present a gatemon based on a Ge/SiGe heterostructure where the Ge hole gas shows superconductivity induced by an Al layer evaporated on top of the buffer layer.}
%with reported relaxation times of up to $2 \, \mu\textrm{s}$~\cite{Casparis2018} and approximately $100 \, \textrm{ns}$~\cite{strickland2023characterizing}. 
The gatemon resonant frequency is electrically tunable in a range of $\sim 5\, \textrm{GHz}$ and exhibits a quasi-monotonic, linear dependence on the gate voltage. 
%(beneficial for the integration in superconducting circuits.) \\
The qubit relaxation time $T_1$ spans from $\sim 80 \, \textrm{ns}$ to $\sim \textrm{20}\, \textrm{ns}$ by varying the gate voltage, while $T_2^*$ does not show any clear trend.
A control qubit on the same material stack using a superconductor-insulator-superconductor (SIS) junction exhibits a $T_1$ about five times longer than the semiconductor junction, and $T_2^*$ approaches $ 2 \, T_1$.
We discuss potential limitations in the Ge gatemon arising from the substrate and device layout.
%Casparis et. al.~\cite{Casparis2018} demonstrated improved gatemon performance on InAs/InGaAs heterostructures by eliminating the buffer and oxide layer, significantly reducing microwave losses. On the other hand, our Ge gatemon device, similar to the approach in Strickland et. al.~\cite{strickland2023characterizing}, retained both the buffer and oxide layer. In this paper, we establish that our Ge gatemon device exhibits comparable performance to that reported in Ref.~\cite{strickland2023characterizing}. \\

\section{Device}

An optical image of our device is shown in Fig.~\ref{fig:device}. We adopt a planar 'Xmon' geometry, known to maintain balance among coherence, connectivity, and swift control~\cite{PhysRevLett.111.080502,Barends2014}.
The core of the device is the semiconductor junction defined on the U-shaped mesa (see Fig.~\ref{fig:device}b and c) with a width of $\sim 450 \, \textrm{nm}$ and a junction length of $\sim 150\, \textrm{nm}$.
The critical current $I_c$
through the semiconductor junction determines the Josephson energy ($E_{\textrm{J}} = \hbar I_{\textrm{c}} / 2e$, with $\hbar$ the reduced Planck constant and $e$ the electron charge), which sets the qubit frequency together with the charging energy $E_{\textrm{c}}$ of the cross-shaped island.
The island and the semiconductor junction together form the qubit circuit. The charging energy of the island is designed to be $E_{\textrm{c}}/ h  \sim 200\,\textrm{MHz}$ to set the qubit frequency to be compatible with the $2-10 \, \textrm{GHz}$ range of our electronics and to operate in the transmon regime $E_{J} \gg E_{C}$~\cite{PhysRevA.76.042319}. The gatemon is coupled to a notch-type, $\lambda/4$ resonator with a loaded quality factor of $Q_{\textrm{l}} \sim 1400$ (see Fig.~\ref{fig:gextract}b). The fabrication process flow is detailed in the Methods Section.

\section{Resonator Spectroscopy}
We investigate the gate tunable qubit-resonator interaction by monitoring the feedline scaterring parameter $S_{21}$. The microwave power is kept low enough such that the average photon population of the resonator is approximately 1.  
By applying a voltage $V_{\textrm{g}}$ on the Ti/Pd gate, we modify the critical current of the junction, which turns into a modulation of the qubit frequency: $f_{\textrm{q}} \propto \sqrt{I_{\textrm{c}} \left(V_{\textrm{g}} \right)}$.
%By applying voltage on the Ti/Pd gate, we tune the critical current of the junction, which in turn tunes the qubit frequency $f_{q} \propto \sqrt{I_{c} \left(V_{g} \right)}$.
As displayed in Fig.~\ref{fig:device}e, the qubit exhibits one pronounced avoided crossing with the resonator.
When the qubit is in resonance with the resonator, we observe two peaks in $|S_{21}|$ (at frequency $f_{+}$ and $f_{-}$, see Fig.~\ref{fig:device}f), a hallmark of two hybridized qubit-resonator states due to the strong coupling regime. 
To extract the coupling strength $g$ we fit the splitting: $\delta = f_{+} - f_{-}$ according to the following equation~\cite{PhysRevLett.115.127001}:
\begin{equation}
    \delta = \sqrt{\left(f_{\textrm{q}}-f_{\textrm{c}} \right)^{2} +4\left(g/2\pi\right)^{2}},
    \label{eq:one}
\end{equation}
as illustrated in Fig.~\ref{fig:gextract}c, where $f_{\textrm{q}}$ and $f_{\textrm{c}}$ are the bare qubit and resonator frequencies respectively.
The fit yields $g / 2\pi \approx 270 \, \textrm{MHz}$, consistent with electrostatic simulations.
We also extract the qubit frequency: 
\begin{equation}
 f_{\textrm{q}} = \left(f_{+}+f_{-} \right)-f_{\textrm{c}},
 \label{eq:two}
\end{equation}
shown with a red dashed line in Fig.~\ref{fig:device}f.
We obtain the bare resonator frequency from an independent high-power resonator spectroscopy measurement. 
We note that, in contrast to nanowire gatemons~\cite{PhysRevApplied.15.054001,PhysRevLett.125.056801}, we observe only one avoided crossing (refer to Fig.~\ref{fig:gextract} for a broader gate voltage range), suggesting a smooth and monotonic (albeit still suffering from discrete charge jumps) frequency dispersion.
In Fig.~\ref{fig:seconddeviceonetone}, we present additional resonator spectroscopy on a separate device with $\sim 50 \, \textrm{nm}$ shorter channel length exhibiting similar behaviour.

%%% Figure 2 %%%%
\begin{figure*}[!htbp]
\includegraphics[width=\textwidth]{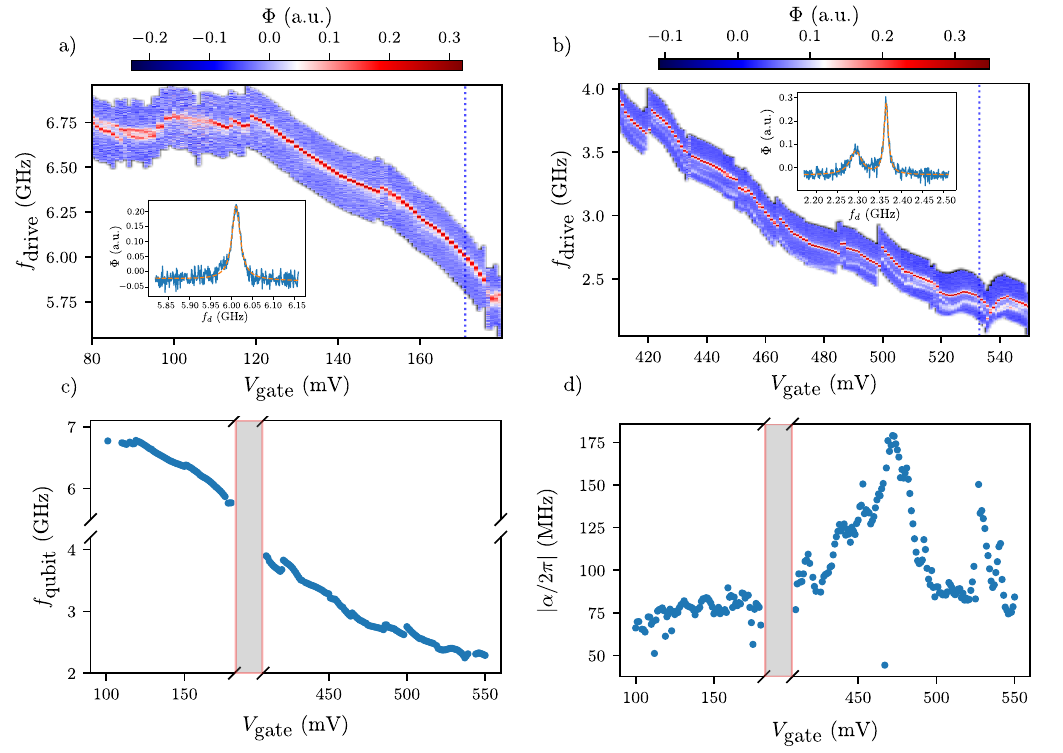}% Here is how to import EPS art
\caption{\label{fig:twotone} Pulsed qubit spectroscopy.
a)-b) Two-tone spectroscopy data after subtracting the average of each column and normalizing the trace.
In panel a (b), the qubit frequency is set above (below) the resonator frequency. We measure the transmitted signal phase after a $2 \, \mu$s excitation. 
In b), we observe the $\ket{1} \rightarrow \ket{2}$ due to a residual excited state population.
The insets show a linecut at $V_{\textrm{g}} = 171 \, \textrm{mV}$ and $V_{\textrm{g}} = 533 \, \textrm{mV}$, respectively, along the blue dashed lines.
The $x$ and $y$ axes correspond to $f_{\textrm{drive}}$ and phase, respectively.
c) Extracted qubit frequency from a) and b). We fit each trace with a Lorentzian (a pair of Lorentzian in the case of panel b), and the center yields the qubit frequency. 
d) Extracted anharmonicity from three-tone spectroscopy as explained in the main text. The anharmonicity is reduced compared to $-E_{\textrm{c}}$, indicating a non-sinusoidal current-phase relation.}
\end{figure*}
%%% Figure 2 %%%%

\section{Qubit spectroscopy}
\begin{figure*}[!htbp]
    \centering
    \includegraphics[width=\textwidth]{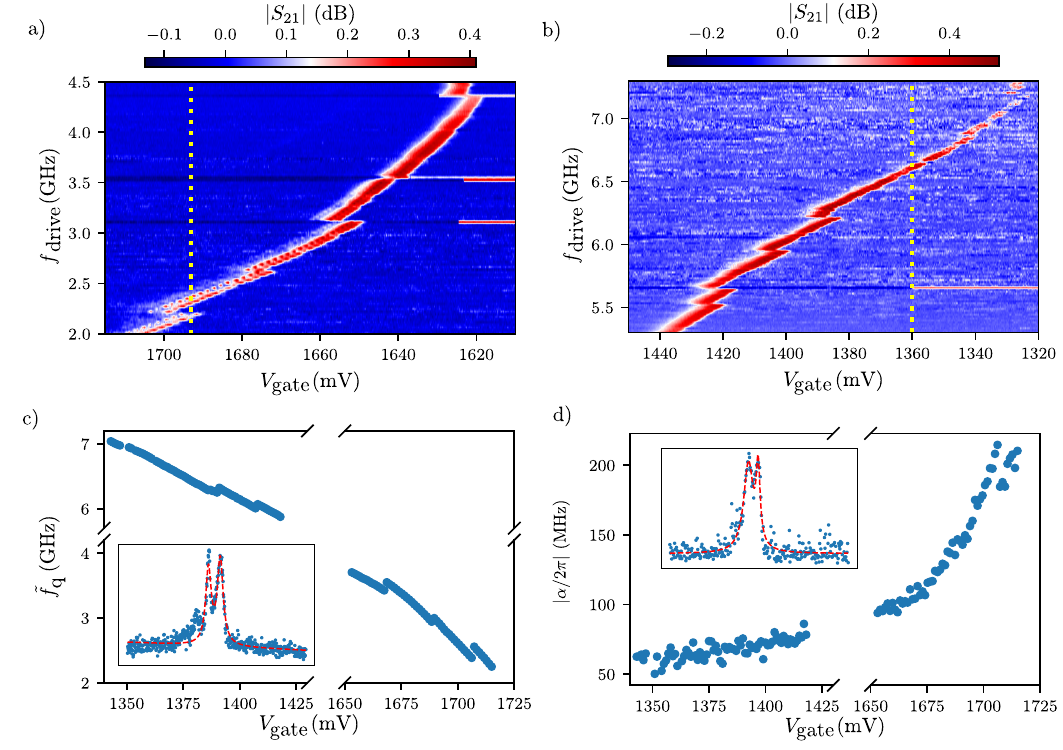}
         \caption{Continuous-wave two-tone spectroscopy on a second gatemon device. The simplicity of continuous measurements allowed for quick characterization. a)-b) Raw two-tone spectroscopy data acquired with a VNA. We choose a high enough power to see the $\ket{0} \rightarrow \ket{2}$ transition via a two-photon process to obtain the anharmonicity with the same measurement. The qubit broadens due to photon shot noise as it moves closer to the resonator since the measurement and drive tone are applied simultaneously. We have subtracted the average of each column at every gate voltage. c) Extracted qubit frequencies versus gate voltage. The inset shows a linecut at $V_{g} = 1693\, \textrm{mV}$ indicated by the yellow dashed line in panel a). We fit each trace with a double Lorentzian curve. The center of the second peak yields the qubit frequency. d) Extracted anharmonicity versus gate voltage. The inset shows a linecut at $V_{\textrm{g}} = 1360\, \textrm{mV}$ indicated by the yellow dashed line in panel b). The distance between the peaks is the half of the anharmonicity.} 
    \label{fig:seconddevicetwotone}
\end{figure*}

\begin{figure*}[!htbp]
    \centering
    \includegraphics[width=\textwidth]{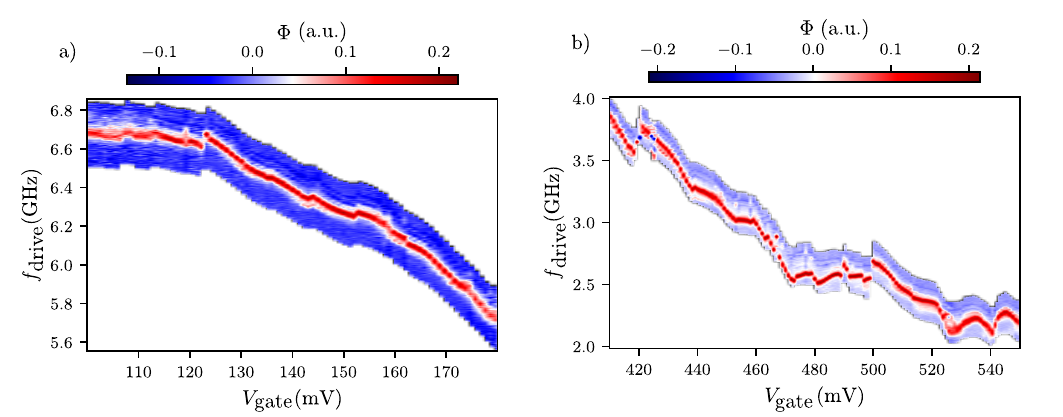}
    \caption{Three-tone spectroscopy measurements used to extract the anharmonicity presented in Fig.~\ref{fig:twotone}d, with $f_{\textrm{q}} > f_{\textrm{r}}$ and $f_{\textrm{q}} < f_{\textrm{r}}$ shown in panel a) and b) respectively.  The position of the measured peak yields the frequency of the $\ket{1} \rightarrow \ket{2}$ transition. At $V_{\textrm{g}} < 100 \, \textrm{mV}$, we observe two peaks in the two-tone spectroscopy measurement. Thus, we have left that region out in that measurement. }
    \label{fig:anharmonicityraw}
\end{figure*}

We now investigate the qubit frequency dependence in a broader gate voltage range than Fig.~\ref{fig:device}f.
To do that, we move the qubit to the dispersive regime ($|\Delta| = |f_{\textrm{q}}-f_{\textrm{r}}| > g/2\pi$) to probe the qubit eigenstates. We adopt a conventional two-tone spectroscopy technique by applying a $2\, \mu\textrm{s}$ long drive tone on the gate line followed by a $150 \, \textrm{ns}$ readout pulse on the resonator line.
When the qubit drive matches $f_{\textrm{q}}$, we observe a peak in the resonator phase response.
We choose the measurement frequency at each gate voltage to obtain all information in the phase of the measured signal.
The observed peaks in Fig.~\ref{fig:twotone}a and b shift consistently with gate voltage; therefore, we attribute it to the qubit mode. The critical current and, thus, the qubit frequency increases with decreasing gate voltage since the charge carriers are holes.
In Fig.~\ref{fig:twotone}b, a second, faint line appears, corresponding to the $\ket{1} \rightarrow \ket{2}$ transition, which is a signature of a residual thermal population of the excited state.
We do not observe this transition when the qubit is above the resonator. Similarly to the resonator spectroscopy, we see a monotonic gate voltage dependence of the qubit frequency interrupted by discontinuities at specific gate voltages, which we attribute to charge jumps. 
Below $V_{\textrm{g}} < 100 \, \textrm{mV}$ the qubit enters an unstable regime, exhibiting two peaks in two-tone spectroscopy likely due to a two-level fluctuator interacting directly with the qubit~\cite{PhysRevLett.120.100502}. Therefore, we disregard the data at $V_{\textrm{g}} < 100\, \textrm{mV}$ in the following analysis.\\
The qubit frequency can be tuned over a range spanning a few GHz, as shown in Fig.~\ref{fig:twotone}c. The junction is tunable even in a broader range, but the diminished readout visibility did not allow us to probe the qubit outside the $2-7 \, \textrm{GHz}$ range. The grey-shaded areas in Fig.~\ref{fig:twotone} panels c and d indicate the strong coupling regime discussed in Fig.~\ref{fig:device}f and parts of the dispersive regime where two-tone spectroscopy is not possible because of the reduced readout visibility.\\
%In Fig.~\ref{fig:twotone}, we plot the extracted qubit frequencies that reveal a tunability in a broad range from $2$ to almost $7$ GHz. 
We extract the anharmonicity at each gate voltage with a separate measurement by using three tones. 
A fixed frequency tone saturates the $\ket{0} \rightarrow \ket{1}$ transition while the frequency of a second tone is swept. We plot the measurement data in Fig.~\ref{fig:anharmonicityraw}, while in Fig.~\ref{fig:twotone}d, we summarize the extracted anharmonicities at all measured gate voltages. All measured anharmonicites are lower than $-E_{\textrm{c}} \approx -200\, \textrm{MHz}$, a signature of non-sinusoidal current-phase relation~\cite{PhysRevLett.67.3836}, varying from $-180$ to $-60 \, \textrm{MHz}$.
The variation in anharmonicity illustrates how the gate voltage influences transparency~\cite{PhysRevB.97.060508}, with increased transparency resulting in lower anharmonicity, as seen in Fig.~\ref{fig:twotone}d.
% While reduced anharmonicity speaks of a good quality superconductor-semiconductor interface, since it indicates non-sinusoidal current-phase relation~\cite{PhysRevB.66.184513}, it hinders fast qubit operation.
While reduced anharmonicity indicates a good interface quality between the proximitized Ge and the Ge forming the weak link, %since it indicates non-sinusoidal current-phase relation~\cite{PhysRevB.66.184513}, 
it also hinders fast qubit operations.
% \textcolor{red}{I suggest to reformulate it as: While reduced anharmonicity speaks of a good interface quality between the proximitized Ge and the Ge forming the weak link, it hinders fast qubit operation.}
In the case of a short, ballistic junction, a fourfold decrease in anharmonicity is expected compared to an SIS junction as the transparency reaches one~\cite{PhysRevB.97.060508}. Our device shows an approximately threefold decrease accompanied by pronounced fluctuations around certain gate voltage values. %Following Ref.~\cite{Valentini2024}, we hypothesize that our junction lies in the diffusive regime ($l > l_{\textrm{h}}$, where $l$ is the junction length and $l_h$ is the hole mean free path), explaining the observed fluctuations.
In Fig.~\ref{fig:seconddevicetwotone}, a similar device with the same width but with a $\approx 50\, \textrm{nm}$ shorter junction shows almost monotonic gate voltage dependence; the origin of this difference calls for further investigations.
%further supporting the assumption that the fluctuations in Fig.~\ref{fig:twotone}d are due to the diffusive nature of the junction.

%%%Figure 3%%%
\begin{figure*}[!htbp]
\includegraphics[width=\textwidth]{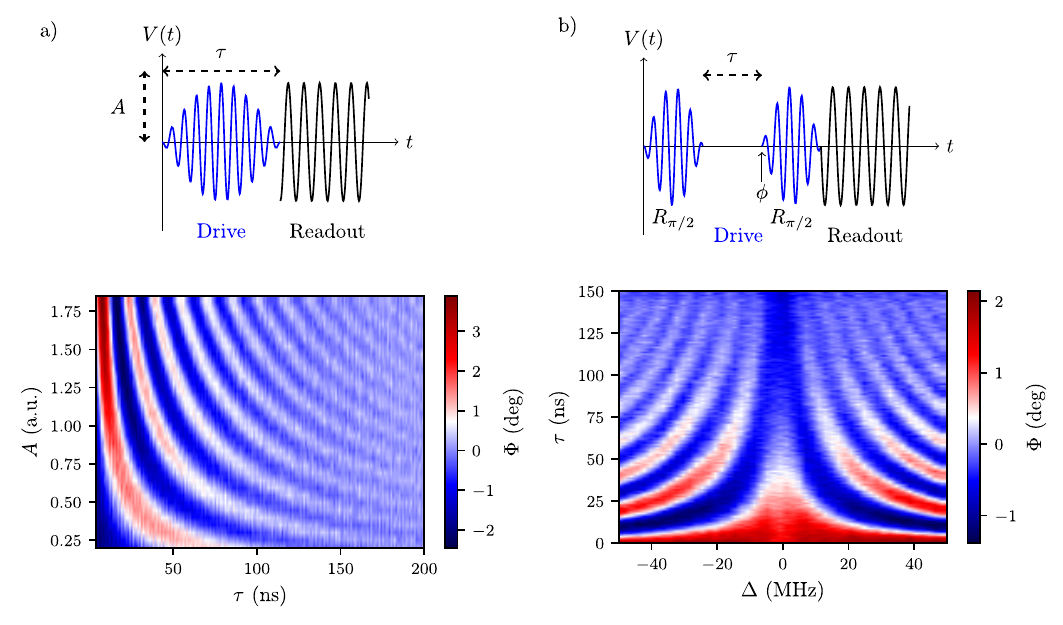}% Here is how to import EPS art
\caption{\label{fig:rabi} Coherent control of the gatemon. a) Rabi oscillations at $f_{\textrm{q}} \approx 3.66 \, \textrm{GHz}.$ We apply a cosine-shaped drive pulse directly on the gate line, followed by a readout pulse on the resonator line.
b) Ramsey fringes with virtual Z gates. We rotate the frame of reference to mimic an extra rotation.
The detuning defines the angle of rotation: $\phi = \Delta \cdot \tau$. In both a) and b), the measured response is averaged over 50000 traces, and the average of each column has been subtracted to account for the slow drift of the readout resonator.}
\end{figure*}
%%% Figure 3 %%%

\section{Coherent control}
Next, we show coherent manipulation of the gatemon states. At a fixed gate voltage, we demonstrate \textit{X}-\textit{Y} rotations by applying a cosine-shaped drive pulse (indicated blue in Fig.~\ref{fig:rabi}a) followed by a readout pulse on the resonator line. We dynamically vary the amplitude and the duration of the drive pulse at each sequence. The plot in Fig.~\ref{fig:rabi}a shows a typical Rabi pattern, featuring oscillations with an increasing frequency as the drive amplitude increases. Notably, the presented data is averaged over an hour, showcasing the stability of the sample at fixed gate voltage on that timescale.\\
After having calibrated the length of our $\pi$ pulse with the previously shown Rabi measurements, we move to \textit{Z} rotations. In Fig.~\ref{fig:rabi}b, we employ virtual $Z$ gates~\cite{PhysRevA.96.022330}, involving two $\pi/2$ half pulses with an offset in the second pulse phase, introducing an artificial detuning $\left( \Delta \right)$. This approach mitigates the limited visibility and unwanted AC Stark shift when using a detuned pulse. We obtain a Ramsey pattern by sweeping the virtual detuning and the idle between the two $\pi/2$ pulses.

%%% Figure 4 %%%
\begin{figure*}[!htbp]
\includegraphics[width=\textwidth]{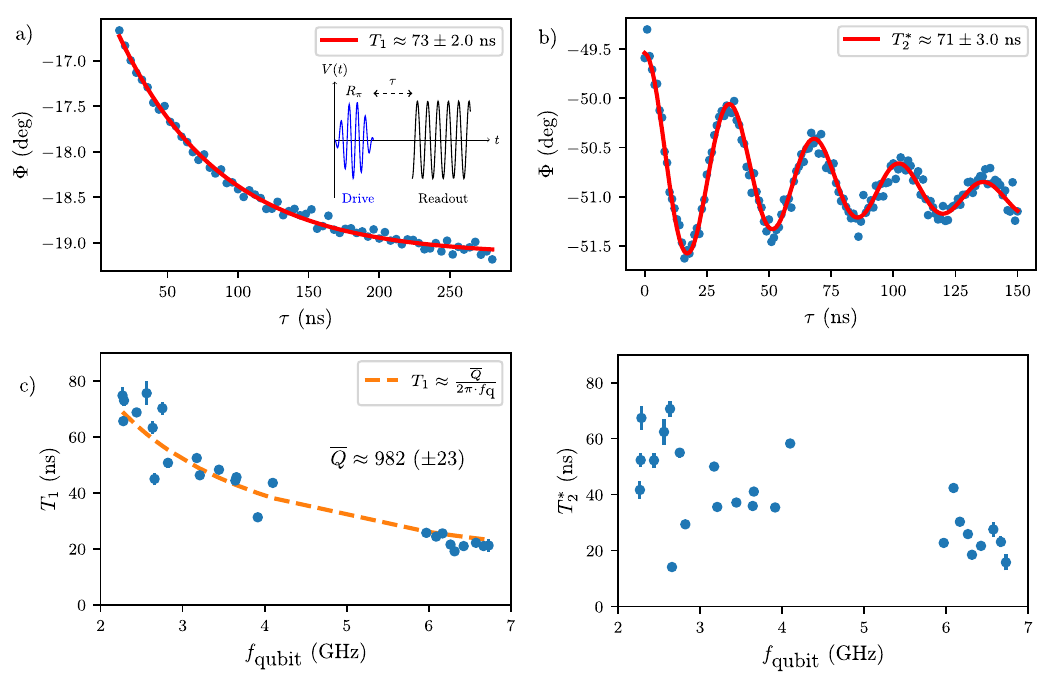}% Here is how to import EPS art
\caption{\label{fig:coherence} Relaxation and coherence time measurements.
a) $T_{1}$ measurement at ${f_{\textrm{q}} \approx 2.8 \, \textrm{GHz}}$. A $10$\, ns $\pi$ pulse brings the qubit to the excited state, followed by a certain waiting time and the readout pulse. The solid red line is a fit to the exponential curve. 
b) $T_{2}^{*}$ measurement at $f_{\textrm{q}} \approx 2.8 \, \textrm{GHz}$. The pulse sequence is identical to the one in Fig.~\ref{fig:rabi}b. The solid curve fits a damped sinusoidal curve on a linear background. The error bars represent the standard deviations of the fit.
c) Relaxation time measurements as a function of qubit frequency.
The error bars represent the standard deviations of the fit, depicted above.
The dashed line indicates a fit to the function shown in the legend.
We extract an effective quality factor lower than the $Q_{i}$s of bare resonators on a similar substrate.
d) Coherence time as a function of qubit frequency. The measured coherence times do not reach the $2T_{1}$ limit.} 
\end{figure*}
%%% Figure 4 %%%%

\section{$T_{1}\; \textrm{and}\; T_{2}^{*}\, \textrm{measurements} $ }
Having calibrated the pulse sequences using previously conducted measurements, we proceed to characterize our qubit relaxation and coherence times.
We determine the relaxation time $T_{1}$ by initializing the qubit in the $\ket{1}$ state and adjusting the waiting time $\tau$ before applying the readout pulse. 
The characteristic decay time of the measured response is $T_{1}$, which is extracted by fitting the data with $A\exp(-\tau/ T_{1})+B$, as illustrated in Fig.~\ref{fig:coherence}a. 
Next, we measure the coherence time, $T_{2}^{*}$, using the same pulse sequence used in the Ramsey plot in Fig.~\ref{fig:rabi}b. We choose an artificial detuning greater than the decoherence rate to observe a sufficient number of oscillations.
This allows us to reliably extract $T_{2}^{*}$ from $A\exp(-\tau/ T_{2}^{*})\cos(2\pi f \tau+\phi)+B+C \tau$, as depicted in Fig.~\ref{fig:coherence}b. 
Following this, we replicate the abovementioned procedures across multiple gate voltages, including qubit spectroscopy, Rabi, and Ramsey measurements.
The results are presented in panels c) and d) of Fig.~\ref{fig:coherence}, where we display the extracted relaxation and coherence times as functions of the qubit frequency. The analysis yields a maximum $T_{1}$ of $\approx 73 \, \textrm{ns}$ across the scanned gate voltage range at $f_{q} \approx 2.8 \, \textrm{GHz}$.
Additionally, we obtain a maximum  $T_{2}^{*}$ of $71 \, \textrm{ns}$ within the explored gate voltage range.

% Coherence times exhibit a non-monotonic trend \textcolor{blue}{where is it non-monotonic?} and lag from the $2T_{1}$ limit. Thus, the coherence of the qubit is not constrained by energy relaxation; instead, it is influenced by dephasing attributed to charge fluctuators within the oxide (which covers the entire chip) or gate voltage noise (originating from the electronics)~\cite{Casparis2018}, with a potential for improvement in these times.

%%% Figure 5 %%%
\begin{figure*}[!htbp]
    \centering
    \includegraphics[width=\textwidth]{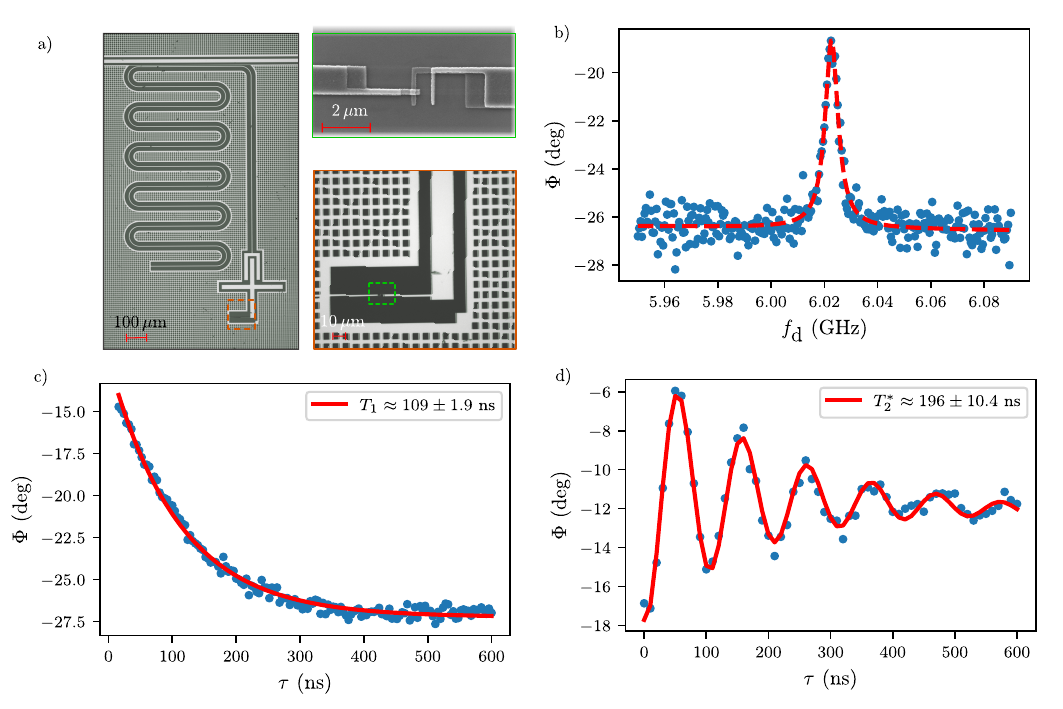}
    \caption{Al transmon characterization on etched SiGe substrate. a) Overview of the device. We replaced the semiconductor junction with a shadow-evaporated Al-$\text{AlO}_x$-Al junction. Otherwise, we kept the device geometry the same.  b) Qubit characterization with pulsed two-tone spectroscopy. The red dashed line represents a fit to a Lorentzian function. We choose the measurement frequency such that all information is contained in the phase of the measured signal.  c). Relaxation time measurement using the same pulse sequence shown in Fig.~\ref{fig:coherence}. The solid red line is a fit to an exponential decay. d) Coherence time measurement using the same pulse sequence as shown in Fig.~\ref{fig:coherence}. The solid red line is a fit to a decaying sinusoidal oscillation.}
    \label{fig:transmon}
\end{figure*}
%%%Figure 5 %%%

%%%Fig 5 Text%%%
\section{Al transmon on SiGe substrate}
% To gain deeper insights, we also compare our device with a fixed-frequency reference transmon measured on the same substrate, shown in Fig.~\ref{fig:transmon} a). The transmon had the same capacitor geometry, and the junction size was also in the same range as the original one. We opted for a fixed-frequency transmon to minimize losses due to relaxation to the gate line.
To gain deeper insights, we compare our device with a fixed-frequency reference transmon measured on a similar substrate and in the same setup, see Fig.~\ref{fig:transmon}a. This transmon shares identical capacitor geometry, and its junction size ($\sim$ 200 nm X 180 nm) falls within the same range as the gatemon device presented ($\sim$ 400 nm X 150 nm). The choice of a fixed-frequency transmon aims to minimize losses in relaxation channels other than dielectric losses in the substrate.
Notably, the hole gas is removed on the entire chip, and the junction is of a conventional SIS type, with Al as superconductor (S) and $\text{AlO}_x$ as an insulator (I).
We determine the qubit frequency and extract $T_{1}$ and $T_{2}^{*}$ times using techniques explained before. The measurements are detailed in Fig.~\ref{fig:transmon} b,c, and d. We find $T_{1} \approx 109 \, \textrm{ns}$ and $T_{2}^{*} \approx 196 \, \textrm{ns}$, very close to the upper limit of $2T_{1}$.

\section{Discussion}
To elucidate the underlying loss mechanisms of our gatemon, we draw comparisons between our findings on the semiconductor transmon and analysis from bare resonators measured on the same substrate, as outlined in Ref.~\cite{Valentini2024}. 
We then complement the discussion by including the Al transmon figures in the comparison  (Fig.~\ref{fig:transmon}).\\
We start by converting the measured relaxation times into quality factors by using the expression ${T_{1} = Q / (2\pi f_{\textrm{q}})}$~\cite{10.1063/5.0017378}. 
In Fig.~\ref{fig:coherence}c, we fit the obtained relaxation times of the gatemon with this equation, which yields an "average" quality factor nearly five times lower than that measured with bare resonators on the same substrates~\cite{Valentini2024}. This suggests additional losses beyond those originating from the substrate.
This conclusion is further supported by the fact that the Al transmon $T_{1}$ is approximately five times higher than in the semiconductor device at the same qubit frequency. We then delve into the disparities between the semiconductor gatemon and the Al transmon/resonators to shed light on the potential loss channels.\\
%\textcolor{blue}{I feel that it would be better to first describe the transmon and then have a compact discussion about what limits us. To shift the text below after the transmon.}
%The observed $T_{1}$ is ca. five times higher than in the semiconductor device at the same qubit frequency, see Fig.~\ref{fig:transmon}c. 
The first significant distinction is the gatemon mesa structure, which features a conducting two-dimensional hole gas covered with Al.
Though a conductive layer normally would cause extra dissipation,  Ref.~\cite{Valentini2024} demonstrated that proximitized germanium (highlighted in red in Fig.~\ref{fig:device}b) does not introduce significant additional losses.
Resonators defined atop proximitized germanium showed comparable internal quality factors (within a factor of two) to those defined with the quantum well etched away. 
Therefore, the presence of proximitized germanium alone cannot solely account for the reduced energy relaxation time.\\
The gatemon differs notably from both the SIS transmon and bare resonators due to the presence of the Pd gate, which may introduce dissipative losses in the gatemon architecture due to its normal metal properties~\cite{PhysRevApplied.19.034021}. On top, the gate itself functions as a coplanar waveguide transmission line, adding another relaxation channel, but we rule out relaxation to the gate line or resonator as a limiting factor (refer to the Supplementary for further details).
Moreover, Atomic Layer Deposition grown $\text{AlO}_x$ is deposited across the entire gatemon device, including the qubit capacitor area where the electric field is concentrated, which can lead to additional dielectric losses~\cite{strickland2023characterizing}. It is also worth mentioning that semiconductor junctions may exhibit subgap states, potentially leading to quasiparticle losses due to external radiation~\cite{PhysRevApplied.15.054001,PhysRevLett.132.017001}.
Finally, the analysis of coherence times of Fig.~\ref{fig:coherence}d reveals a non-monotonic trend, diverging from the $2T_{1}$ limit. 
This observation indicates that the coherence of the qubit is not fully constrained by energy relaxation but also influenced by dephasing attributed to charge fluctuators within the oxide (covering the entire chip)~\cite{Casparis2018}. 
%suggesting room for improvement.
% Additional potential loss mechanisms include relaxation to the drive line or the resonator (discussed in Supplementary) and quasiparticle poisoning\cite{PhysRevLett.132.017001}.
% One obvious culprit is the Pd gate that introduces resistive losses as being a normal metal~\cite{PhysRevApplied.19.034021} and Ref.~\cite{strickland2023characterizing} reports $AlO_{x}$ as a limiting factor in their devices. Other loss mechanisms could be relaxation to the drive line or the resonator and quasiparticle poisoning~\cite{PhysRevLett.132.017001}.
 
%%%Fig % Text%%%

%%%Summary%%%
\section{Summary}

In summary, we have demonstrated a gate-tunable superconductor-semiconductor qubit on planar Ge. Our qubit shows broadband tunability and quasi-linear frequency dispersion, rendering it a promising element for ASQs and superconducting circuits requiring tunable elements. We have characterized the energy relaxation and coherence times, pinpointing the possible limitations with identified areas for improvement. One natural upgrade is to replace the Ti/Pd with either Al, Nb, or Ta, as exemplified in Ref.~\cite{PhysRevApplied.19.034021}. Losses in the oxide can be mitigated by lifting or etching the oxide with an additional fabrication step or switching to a low-loss dielectric, e.g., hBN~\cite{Wang2022}. To prolong the lifetimes of our gatemons beyond the constraints imposed by dielectric losses in the buffer, one can explore strategies such as deep reactive ion etching~\cite{10.1063/1.4907935} and flip-chip technology~\cite{hinderling2024direct,PhysRevApplied.20.034056}.\\
%\textcolor{red}{I should suggest we rephrase as: As Germanium has already solidified its position as a prominent player in the spin qubit domain ~\cite{Borsoi2024,Hendrickx2021} it is naturally considered an alternative platform for ASQs since current experiments are limited by dephasing due to nuclei spins~\cite{pitavidal2023strong}. } 

With germanium already having solidified its position as a prominent player in the spin qubit domain~\cite{Borsoi2024,Hendrickx2021,wang2024operating}, it is naturally considered an alternative platform for ASQs since current experiments are limited by dephasing due to nuclear spins~\cite{pitavidal2023strong}.
To compete with III-V systems in ASQ platforms, the potential of Ge still needs to be demonstrated in the field of superconducting qubits.
In that context, our work constitutes the entry of planar Ge in the field of hybrid qubits. Integrating gatemons in ASQ circuits will loosen up readout requirements by allowing in-situ frequency tuning and capacitive coupling to resonators~\cite{PRXQuantum.3.030311}.
Future challenges will involve addressing issues introduced by the shallow quantum wells~\cite{Valentini2024} and improving the microwave properties of the substrates~\cite{10.1063/5.0038087}.

% Directly before the submission of our manuscript, we became aware of another work dealing with hybrid quantum devices based on planar Ge~\cite{hinderling2024direct}. \textcolor{red}{Why do you feel that we need to write this sentence? Your statement: In that context, our work constitutes the entry of planar Ge in the field of hybrid qubits is still valid, right?}

%%%Summary%%%

\section{Methods}
\subsection{\label{section:fab} Sample fabrication}
First, we define the mesa structure with a depth of $\sim 60\, \textrm{nm}$ by etching away the hole gas with an SF$_{6}$-O$_{2}$-CHF$_{3}$ reactive ion etching process. The hole gas is etched away everywhere, except the mesa area, to remove the conducting hole gas below the microwave circuit. The Josephson junctions (JJs) are formed by evaporating at room temperature $80\, \textrm{nm}$ Al on the mesa after a  $15 \, \textrm{s}$ buffered HF dip. The Al is oxidized for $2$ min at $10$ mbar pressure before venting the evaporation chamber. The microwave circuitry is defined simultaneously with the JJ to reduce the number of fabrication steps. Finally, a plasma-assisted aluminum oxide, approximately $15 \, \textrm{nm}$ thick, is deposited at $150 \, \textrm{°C}$, followed by the evaporation of Ti/Pd gates, where Ti was used as an adhesion layer.
 The thickness of the Al and Pd were chosen to be $20 \, \textrm{nm}$ greater than the mesa depth to ensure proper climbing of the edge.
\subsection{Measurements}
\begin{figure*}[!htbp]
    \centering
    \includegraphics[width=\textwidth]{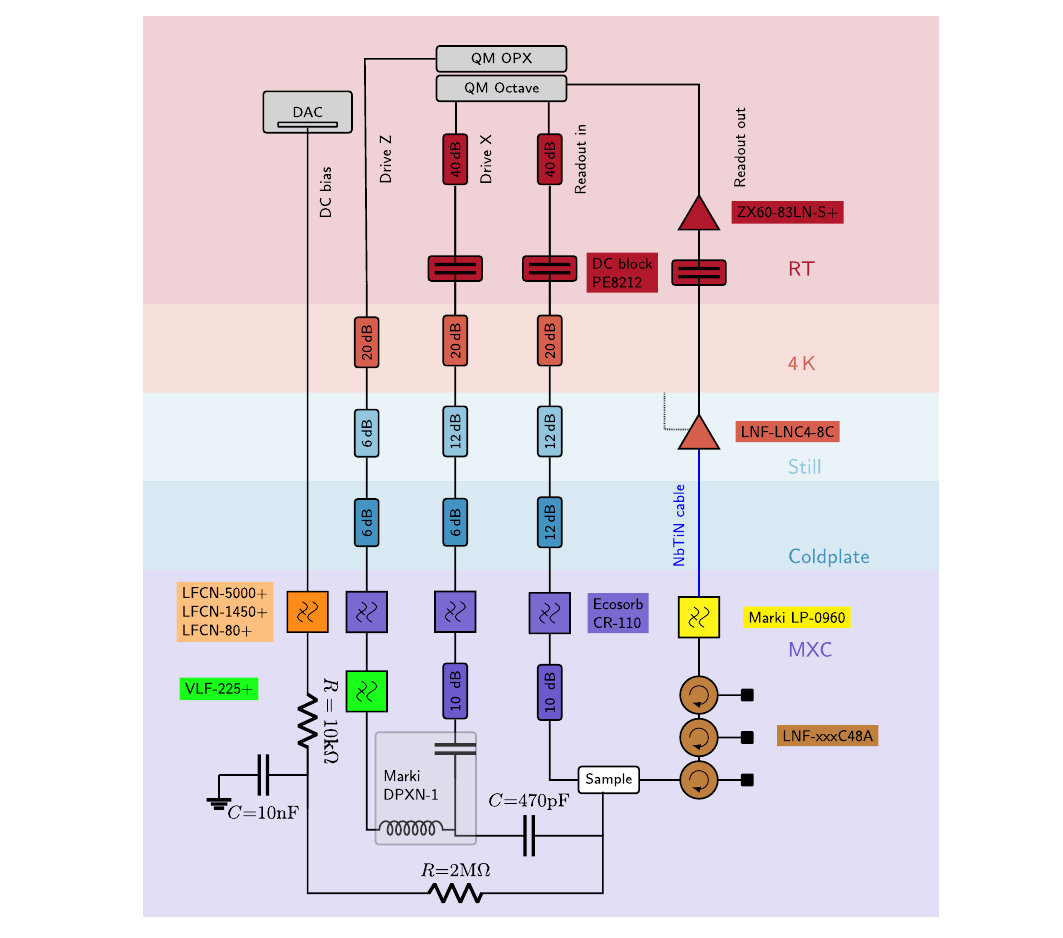}
    \caption{Schematics of the fridge wiring and measurement setup. The \textit{Readout in} and \textit{Readout out} lines are connected to a VNA for continuous-wave measurements. For time-domain measurements, we have used Quantum Machines (QM) instruments. We have used \textit{Drive X} line to send microwave (GHz) pulses, whereas the \textit{Drive Z} line is intended for sending DC pulses. We combine the two lines at the MXC. The \textit{Drive Z} line is not used in this experiment.} 
    %\textcolor{red}{did you intend to stop here? were combined and reached the sample?}}
    \label{fig:meassetup}
\end{figure*}

We performed all measurements in a cryogen-free dilution refrigerator with a base temperature below $10$ mK. The sample was mounted on a custom printed circuit board thermally anchored to the mixing chamber of the cryostat, and electrical connections were made via wire bonding. The schematic of our measurement setup is shown in Fig.~\ref{fig:meassetup}.

\section{Acknowledgements}
We acknowledge Lucas Casparis, Jeroen Danon, Valla Fatemi, Morten Kjaergard and Javad Shabani for their valuable insights and comments. 
This research was supported by the Scientific Service Units of ISTA through resources provided by the MIBA Machine Shop and the Nanofabrication facility. 
This research and related results were made possible with the support of the NOMIS Foundation and the FWF Projects with DOI:10.55776/I5060 and DOI:10.55776/P36507. We also acknowledge the NextGenerationEU PRIN project 2022A8CJP3 (GAMESQUAD) for partial financial support.

\section{Author contribution}
O.S. and L.B. were responsible for fabricating the devices, with O.S. conducting measurements and analyzing the data under the guidance of G.K.
M.V. and O.S. collaborated on developing the nanofabrication recipe for hybrid devices, while M.J. and O.S. were involved in developing the microwave technology for the Ge/SiGe heterostructures.
G.F. contributed to device fabrication and measurements. 
S.C., D.C., and G.I.  were responsible for the growth of the Ge QW.
F.H. and L.K. provided assistance during the experiments. 
F.H., L.K., and J.F. provided the transmon nanofabrication recipe.
O.S., A.C., and G.K. discussed the results and wrote the manuscript with contributions and feedback from all co-authors.

\newpage
\bibliography{apssamp}% Produces the bibliography via BibTeX.

\newpage

\section{Supplementary Data}

\subsection{Additional loss mechanisms}
\begin{figure*}[!htbp]
    \centering
    \includegraphics[width=\textwidth]{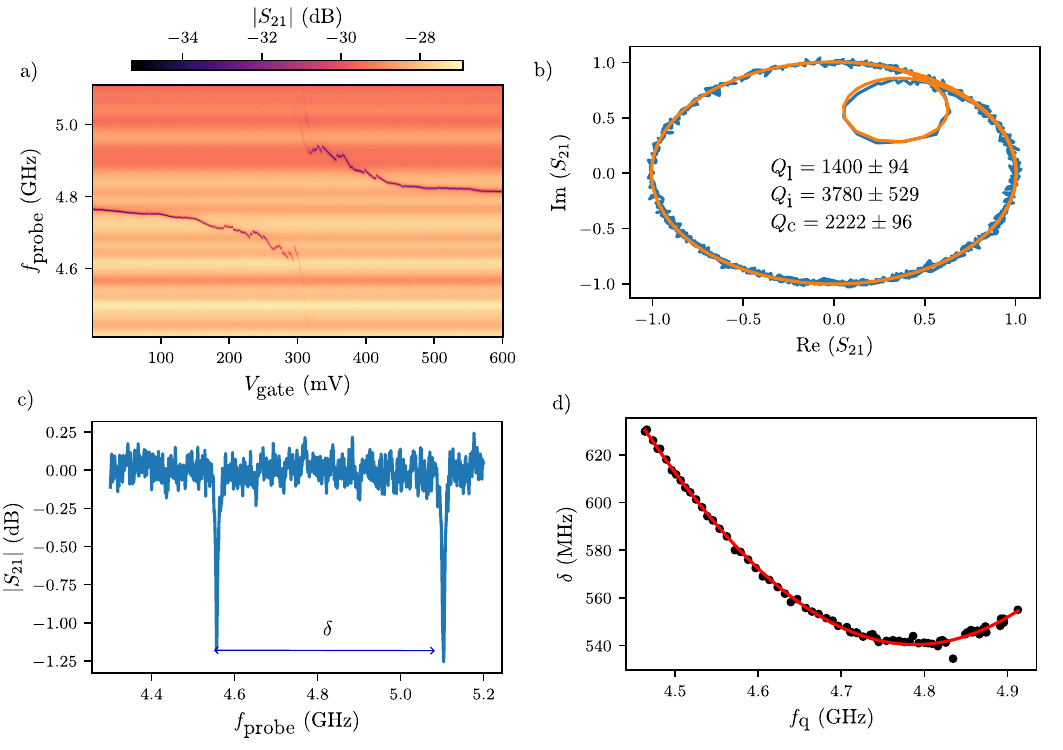}
    \caption{Extended resonator spectroscopy data on the device presented in the main text. a) VNA sweep for a broader gate range without background correction at $n_{\textrm{ph}} \approx 1$ power level. b) Fit to extract the bare resonator properties taken at $V_{\textrm{g}} = 700 \, \textrm{mV}$ where the qubit is very far detuned from the resonator. We fit the complex $S_{21}$ parameter with the \textit{resonator-tools}~\cite{10.1063/1.4907935} package. c) Line-cut from Fig.~\ref{fig:device}e at $V_{\textrm{g}} = 307 \, \textrm{mV}$. We denote the splitting of the hybridized resonator-qubit states by $\delta$ as mentioned in the main text. d) Extracted splitting as a function of the qubit frequency. The solid red line is a fit according to Eq.~\ref{eq:one} using $g$ as a free parameter. }
    \label{fig:gextract}
\end{figure*}

\begin{figure*}[!htbp]
    \centering
    \includegraphics[width=\textwidth]{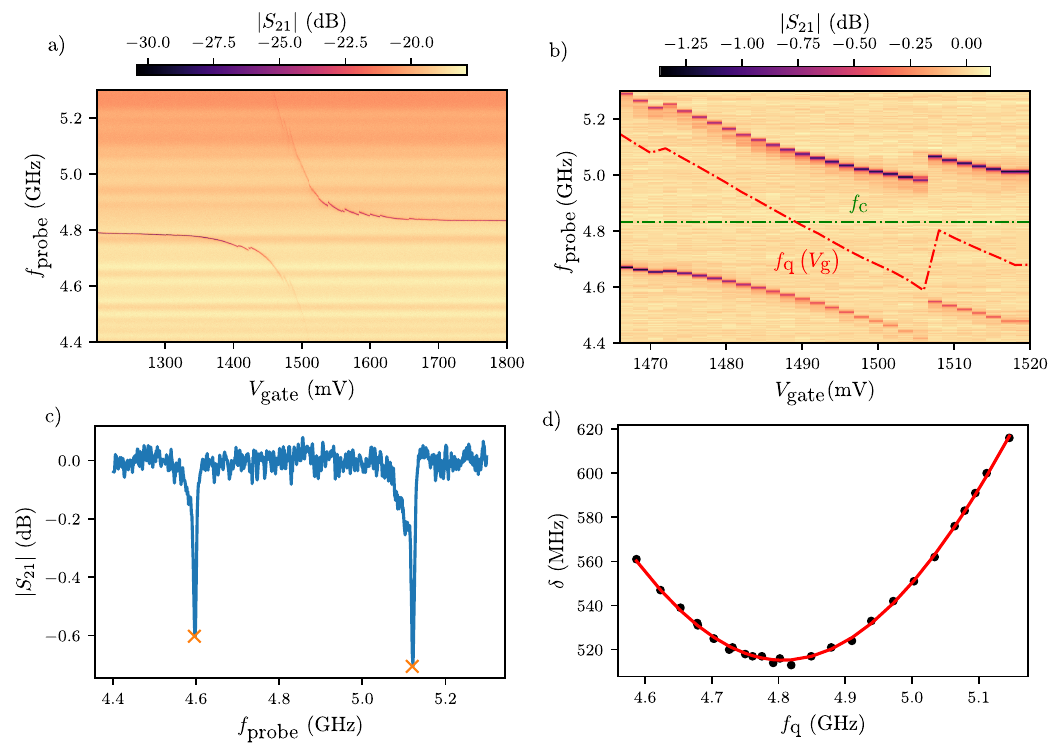}
    \caption{Resonator spectroscopy on a second Ge gatemon device. Apart from the junction length, which is $50 \, \textrm{nm}$ shorter, the second device is nominally identical to the first one discussed in the main text. a) VNA sweep at low photon number. Since the junction is shorter compared to the sample of the main text, we needed to operate in a different gate voltage range to obtain the same critical current. b) Zoom-in of panel a). We observe two peaks due to qubit-resonator hybridization. c). One-dimensional line-cut at $V_{\textrm{g}} = 1494 \, \textrm{mV}$. d) Coupling strength extraction in the same way as demonstrated in Fig.~\ref{fig:gextract}. We extract a coupling value of $g/ 2\pi \approx 257\,\textrm{MHz}$, consistent with $g$ obtained in Fig.~\ref{fig:gextract}d. }
    \label{fig:seconddeviceonetone}
\end{figure*}

\begin{figure*}[!htbp]
    \centering
    \includegraphics[width=0.5\textwidth]{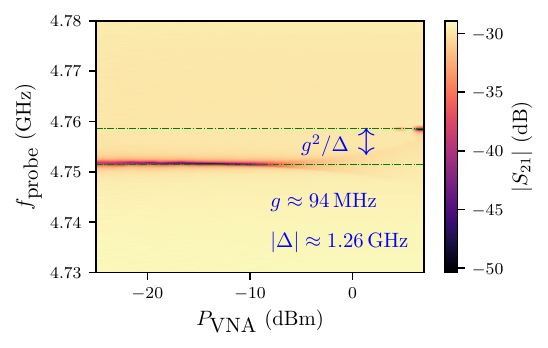}
    \caption{Normalized feed line transmission of the reference transmon as a function of readout power. We observe a Lamb shift $\chi_{0}=g^{2}/\Delta \approx 7\, \textrm{MHz}$ as a signature of coupling to a non-harmonic system. We extract the linewidth $\kappa$ of the resonator at $P_{\textrm{VNA}} = -25 \, \textrm{dBm}.$}
    \label{fig:transmong}
\end{figure*}

The measured $T_{1}$ times range from $75$ to $20$ ns. 
Here, we demonstrate that we are not limited by design constraints, i.e., decay to the resonator and the gate (drive) line. 
To estimate the Purcell limit, we extract the bare resonator properties as depicted in Fig.~\ref{fig:gextract}. 
The obtained Purcell limit at the smallest detuning reads~\cite{PhysRevLett.101.080502}: $T_{\textrm{Purcell}} = \left(\kappa \left(g / \Delta\right)^{2}\right) ^{-1} \approx 400 \, \textrm{ns} $ ($\kappa \approx 2\pi \cdot 3.4 \, \textrm{MHz}$ and $\textrm{max} \left( g / \Delta  \right) \approx 0.34, |\Delta_{\textrm{min}}|\approx 0.8 \, \textrm{GHz}$). 
%\textcolor{red}{If $\kappa$ is a frequency, then we have to put a -1 on the exponent of the 400 ns. I'd add a reference too, maybe PRL 112, 190504 (2014)?} 
The calculated Purcell limit is an order of magnitude higher than the measured $T_{1}$ times. Thus, it is not a limiting factor. 
The same analysis holds for the reference transmons. The coupling strength is derived from a power-dependent resonator spectroscopy measurement, shown in Fig.~\ref{fig:transmong}.  
Our analysis yields a coupling strength of approximately $g / 2 \pi \approx 94 
\,\textrm{MHz}$ , a detuning of approximately $|\Delta| \approx 1.26 
\, \textrm{GHz}$, and a linewidth of approximately $\kappa \approx 2.23 \cdot 2\pi 
\, \textrm{MHz}$. These parameters result in a Purcell time ($T_{\textrm{Purcell}}$) of approximately $12 \, \mu \textrm{s}$, which is two orders of magnitude higher than the measured $T_{1}$.\\
To estimate the relaxation to the drive line, we model the gatemon as an $LC$ resonator capacitively coupled to the unfiltered $Z = 50 \, \Omega$ designed (impedance of the drive line) transmission line, acting as a gate. The environmental $50 \, \Omega$ impedance will dissipate the energy escaping the gatemon circuit. We estimate the $T_{1}$ limit following the formula in Refs.~\cite{relaxation,PhysRevA.76.042319}:
\begin{equation}
    T_{\textrm{drive}} \approx \frac{4C}{RC_{\textrm{g}}^{2}\omega^{2}} \approx 22\, \mu\textrm{s},
\end{equation}
where $Z_{0} \approx 50 \, \Omega$ is the impedance of the feedline, $C \approx 85 \, \textrm{fF}$ is the impedance of the gatemon circuit, $\omega \approx 2 \pi \cdot 5 \, \textrm{GHz} $ is the resonant frequency and $C_{\textrm{g}} \approx 0.57 \, \textrm{fF}$ is the capacitance between the gatemon circuit and the feedline.
%%% Supp. Fig. 1 %%%
%%%Sup. Fig. 1 %%%

\begin{figure*}[!htbp]
    \centering
    \includegraphics[width=\textwidth]{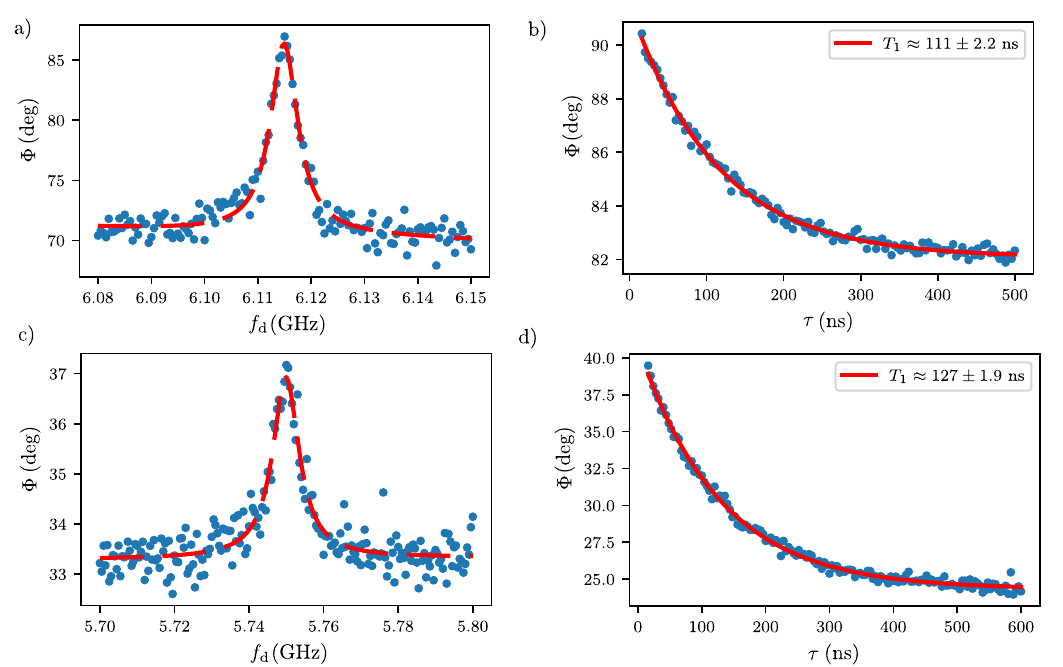}
    \caption{Data from two additional Al transmons, one related to panels a) and b), the other to panels c) and d).  The devices presented are identical to the one shown in Fig.~\ref{fig:transmon}, except for the junction size, which varies to get different qubit frequencies. a), c) Qubit spectroscopy with pulsed two-tone measurement. b), d) Energy relaxation measurements. The measured results are consistent with the ones presented in the main text.}
    \label{fig:transmonextended}
\end{figure*}

\begin{figure*}[!htbp]
    \centering
    \includegraphics[width=\textwidth]{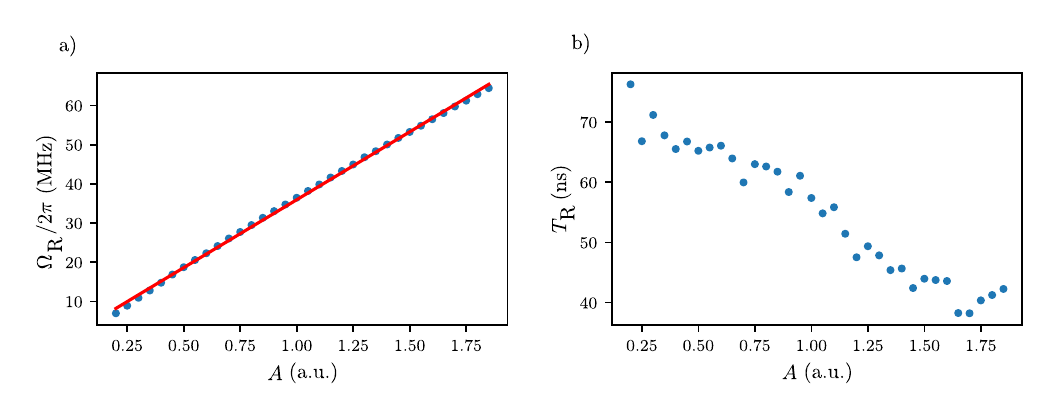}
    \caption{Rabi frequency (panel a) and Rabi times (panel b) as a function of amplitude extracted from Fig.~\ref{fig:rabi} from the main text. At each amplitude, we fitted
the trace with the following expression: $A\cos(\Omega_{\textrm{R}}t+ \varphi)\exp(-t/ T_{\textrm{R}}) + B$. }
    \label{fig:rabisupplementary}
\end{figure*}

\end{document}